\documentclass[journal, 10 pt, a4paper]{IEEEtran}

\newtheorem{thm}{Theorem}

\newtheorem{lem}{Lemma}

\usepackage[final]{graphicx}
\usepackage[reqno]{amsmath}
\usepackage{amssymb}
\usepackage{subfig}
\usepackage{epstopdf}
\usepackage{bm}

\begin{document}

\sloppy

\title{Cell Associations that Maximize the Average \\Uplink-Downlink Degrees of Freedom}
\author{\IEEEauthorblockN{Aly El Gamal\\}
 \IEEEauthorblockA{ECE Department, Purdue University\\ Email: \{elgamala\}@purdue.edu}}

\maketitle

\begin{abstract}
We study the problem of associating mobile terminals to base stations in a linear interference network, with the goal of maximizing the average rate achieved over both the uplink and downlink sessions. The cell association decision is made at a centralized cloud level, with access to global network topology information. More specifically, given the constraint that each mobile terminal can be associated to a maximum of $N_c$ base stations at once, we characterize the maximum achievable pre-log factor (degrees of freedom) and the corresponding cell association pattern. Interestingly, the result indicates that for the case where $N_c \geq 2$, the optimal cell association guarantees the achievability of the maximum uplink rate even when optimizing for the uplink alone, and for the case where $N_c=1$, the optimal cell association is that of the downlink. Hence, this work draws attention to the question of characterizing network topologies for which the problem can be simplified by optimizing only for the uplink or only for the downlink.
\end{abstract}

\section{Introduction}
As we approach the setting of a new standard for cellular communication, the need appears for new strategies to handle the wireless interference management problem by exploiting new technological advances. Two major developments that are expected to affect future generation cellular networks are cloud computing and enhancement of the backhaul infrastructure. In this work, we are attempting to draw attention to the potential of these developments in increasing the rate of communication in interference networks even when relying on simple coding schemes with minimal delay requirements. We consider the simple connectivity model of linear interference networks~\cite{Wyner}, and model the limit of backhaul capacity by assuming that each mobile terminal can be associated with at most $N_c$ base stations in a $K$-user network. The problem we solve is that of identifying the cell associations and coding schemes that achieve the average degrees of freedom across both downlink and uplink sessions.

There has been a recent interest in analyzing cloud based schemes for interference management that assume the availability of cooperative communication~\cite{CRAN}, or multiple antenna systems~\cite{Simeone}. In particular, similar models to the system model considered in this work have been considered in~\cite{ElGamal-Annapureddy-Veeravalli-IT14} but only for the downlink and in~\cite{Ntranos-uplink} but only for the uplink scenario. As for considering a joint framework for both uplink and downlink cooperative communication, a coordinated multi-point interference alignment scheme has been analyzed for fully connected interference networks in~\cite{Annapureddy-ElGamal-Veeravalli-IT11}, where cooperation takes place between transmitters and between receivers. In~\cite{Ntranos}, a duality has been established between a class of zero-forcing interference cancellation schemes in both the uplink and downlink cellular scenarios. In~\cite{Suh}, the degrees of freedom has been analyzed for two multi-antenna base stations that simultaneously serve a set of users in the downlink and another set of users in the uplink.

The key difference in this work from previous attempts is that we identify the optimal cell associations that enable the achievability of the best average rate over both uplink and downlink. In other words, we do not allow the cell association to change between the uplink and downlink sessions. This reflects a practical scenario, where cell association are fixed over both uplink and downlink sessions either to save the overhead or to setup or allocate backhaul links that are required to associate a mobile terminal with multiple base stations. Further, we allow the decision of cell association as well as coding scheme to take into account global information about the network topology, to reflect the cloud application. 

It is worth noting that the considered problem is strongly tied to the problems of interference alignment (see~\cite{Cadambe-IA} and~\cite{Ali-Motahari-Khandani-IT08}) and design of coordinated multi-point coding schemes (see e.g.,~\cite{CoMP-book}). We observe a clear advantage of the coding scheme proposed in this work in Section~\ref{sec:achievability} is that it belongs to the class of one-shot interference cancellation schemes that does not have delay requirements due to symbol extensions as in interference alignment based coordinated multi-point schemes (see e.g.,~\cite{Annapureddy-ElGamal-Veeravalli-IT11}).

The rest of the document is organized as follows. In Section~\ref{sec:problemsetup}, we state the system model. We then start discussing the solution by first recalling previous results in the literature on the downlink scenario in Section~\ref{sec:downlink}. We then show how we solve the average uplink-downlink problem for a simple example in Section~\ref{sec:example}. The main result is introduced and discussed in Section~\ref{sec:main}. Sections~\ref{sec:achievability} and~\ref{sec:converse} provide the proofs of the degrees of freedom inner and upper bounds. We finally provide concluding remarks in Section~\ref{sec:conclusion}. 
\section{Problem Setup}\label{sec:problemsetup}

For each of the downlink and uplink sessions, we use the standard model for the $K-$user interference channel with single-antenna transmitters and receivers,
\begin{equation}
Y_i(t) = \sum_{j=1}^{K} H_{i,j}(t) X_j(t) + Z_i(t),
\end{equation}
where $t$ is the time index, $X_j(t)$ is the transmitted signal of transmitter $j$, $Y_i(t)$ is the received signal at receiver $i$, $Z_i(t)$ is the zero mean unit variance Gaussian noise at receiver $i$, and $H_{i,j}(t)$ is the channel coefficient from transmitter $j$ to receiver $i$ over the time slot $t$. We remove the time index in the rest of the paper for brevity unless it is needed. The signals $Y_i$ and $X_i$ correspond to the receive and transmit signals at the $i^{\textrm{th}}$ base station and mobile terminal in the uplink, respectively, and the $i^{\textrm{th}}$ mobile terminal and base station in the downlink respectively. For consistency of notation, we will always refer to $H_{i,j}$ as the channel coefficient between mobile terminal $i$ and base station $j$.

%
\subsection{Channel Model}\label{sec:channel}
We consider the following linear interference network. The mobile terminal with index $i$ is connected to base stations $i$ and $i-1$, except the first mobile terminal which is connected only to the first base station. More precisely, 

\begin{equation}\label{eq:channel}
H_{i,j} = 0 \text { iff } i \notin \{j,j+1\},\forall i,j \in [K],
\end{equation}
and all non-zero channel coefficients are drawn from a continuous joint distribution. Finally, we assume that global channel state information is available at all mobile terminals and base stations. 

\subsection{Cell Association}
For each $i \in [K]$, let ${\cal C}_i \subseteq [K]$ be the set of base stations for which mobile terminal $i$ is associated with, i.e., those base stations that carry the terminal's message in the downlink and will have its decoded message for the uplink. The transmitters in ${\cal C}_i$ cooperatively transmit the message $M_i$ to mobile terminal $i$ in the downlink. In the uplink, one of the base station receivers in ${\cal C}_i$ will decode $M_i$ and pass it to the remaining receivers in the set. We consider a cell association constraint that bounds the cardinality of the set ${\cal C}_i$ by a number $N_c$; this constraint is one way to capture a limited backhaul capacity constraint where not all messages can be exchanged over the backhaul. 
\begin{equation}\label{eq:backhaul_constraint}
|{\cal C}_i| \leq N_c, \forall i\in [K].
\end{equation}

We would like to stress on the fact that we only allow full messages to be shared over the backhaul. More specifically, splitting messages into parts and sharing them as in~\cite{Wigger}, or sharing of quantized signals as in~\cite{Ntranos} is not allowed. 

\subsection{Degrees of Freedom}
Let $P$ be the average transmit power constraint at each transmitter, and let ${\cal M}_i$ denote the alphabet for message $M_i$. Then the rates $R_i(P) = \frac{\log|{\cal M}_i|}{n}$ are achievable if the decoding error probabilities of all messages can be simultaneously made arbitrarily small for a large enough coding block length $n$, and this holds for almost all channel realizations. The degrees of freedom $d_i, i\in[K],$ are defined as $d_i=\lim_{P \rightarrow \infty} \frac{R_i(P)}{\log P}$. The DoF region ${\cal D}$ is the closure of the set of all achievable DoF tuples. The total number of degrees of freedom ($\eta$) is the maximum value of the sum of the achievable degrees of freedom, $\eta=\max_{\cal D} \sum_{i \in [K]} d_i$.

For a $K$-user channel, we define $\eta(K,N_c)$ as the best achievable $\eta$ on average taken over both downlink and uplink sessions over all choices of transmit sets satisfying the backhaul load constraint in \eqref{eq:backhaul_constraint}. 
In order to simplify our analysis, we define the asymptotic per user DoF $\tau(N_c)$ to measure how $\eta(K,N_c)$ scales with $K$ while all other parameters are fixed,
\begin{equation}
\tau(N_c) = \lim_{K\rightarrow \infty} \frac{\eta(K,N_c)}{K},
\end{equation}

We further define $\tau_D (N_c)$ and $\tau_U (N_c)$ as the asymptotic per user DoF when we optimize only for the downlink and uplink session, respectively.
\section{Optimal Downlink Cell Association}\label{sec:downlink}
In~\cite{ElGamal-Annapureddy-Veeravalli-IT14}, the per user DoF for the downlink $\tau_D (N_c)$ was characterized for all values of $N_c$ as,
\begin{equation}
\tau_D (N_c) = \frac{2N_c}{2N_c +1}.
\end{equation}  
The optimal cell association has a repeated pattern every $2N_c+1$ users, and hence, it suffices to describe it for the first $2N_c+1$ users. The mobile terminal with index $N_c+1$ is not served in the optimal scheme. Note that this is the case as the goal is to maximize the sum degree of freedom. However, fairness can be achieved by shifting the scheduling strategy by one index every resource slot (time or frequency). The remaining users are grouped into the following clusters:

\begin{eqnarray*}
{\cal S}_1 &=& [N_c]
\\{\cal S}_2 &=& \{N_c+2,N_c+3,\ldots,2N_c+1\}
\end{eqnarray*}

and the cell association is given by the following description.

${\cal C}_{i}=
\begin{cases}
\{i,i+1,\ldots,N_c\}, \quad &\forall i \in {\cal S}_1,\\
\{i-1,i-2,\ldots,N_c+1\},\quad &\forall  i \in {\cal S}_2.
\end{cases}$\\

\begin{figure}[htb]
\centering
\includegraphics[width=0.4\columnwidth]{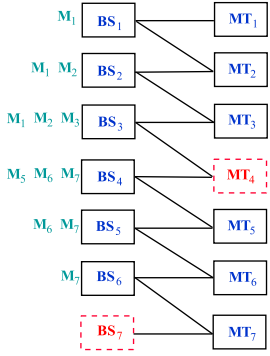}
\caption{Illustration of the cell association that acheives $\tau_D (3)$. The red dashed nodes are inactive.}
\label{fig:ncthreedownlink}
\end{figure}

Figure~\ref{fig:ncthreedownlink} illustrates a description of the optimal downlink cell association for the case when each mobile terminal can be associated with three base stations, i.e., $N_c=3$. In this case, $\frac{6}{7}$ per user DoF is achieved by delivering messages $M_1, M_2, M_3, M_5, M_6, M_7$ through base stations $1, 2, 3, 4, 5, 6$, respectively, and using the knowledge of other messages to cancel undesired interference. Since $6$ messages out of $7$ can be delivered while completely eliminating interference (note that this is possible since channel state information is available at transmitters), then $\frac{6}{7}$ per user DoF is achieved. 

\section{Example: $N_c=2$}\label{sec:example}
\begin{figure}
  \centering
\subfloat[]{\label{fig:nctwodownlink}\includegraphics[height=0.2\textwidth]{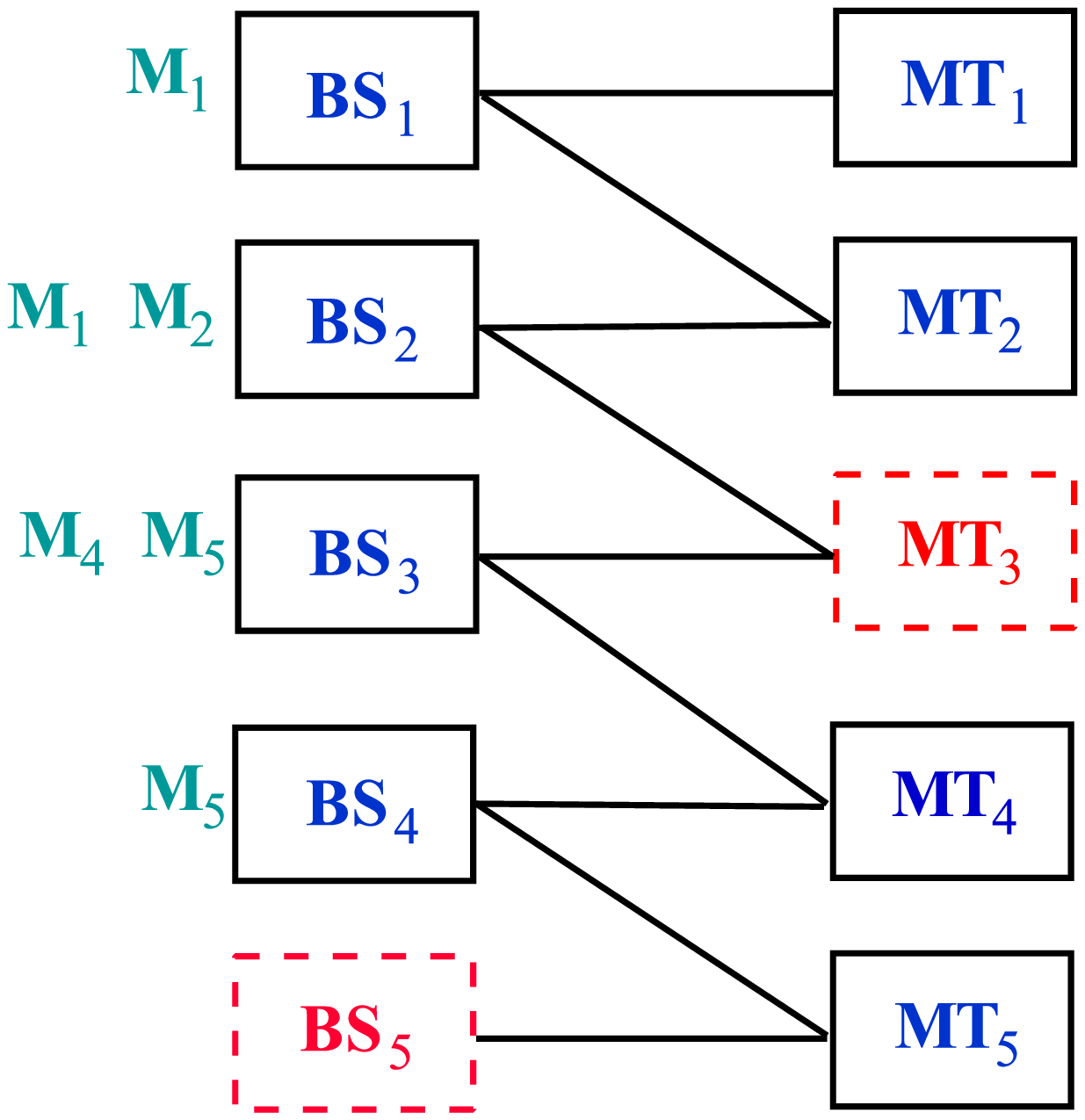}}                
\quad\quad\quad\quad\subfloat[]{\label{fig:nctwouplink}\includegraphics[width=0.16\textwidth]{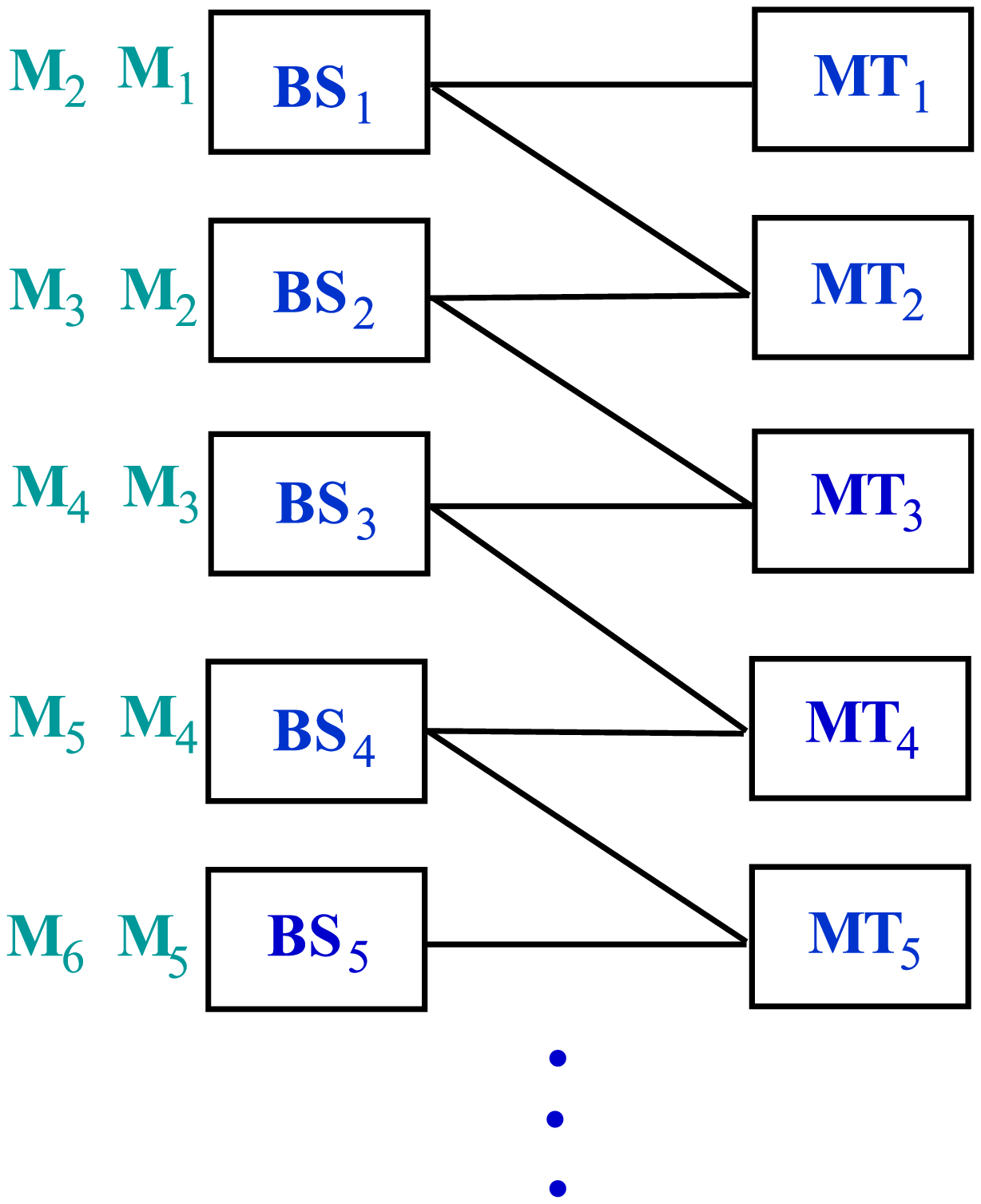}}
  \caption{Figure showing optimal cell associations for the case of $N_c=2$. In $(a)$, the optimal cell association for the downlink is shown. In ($b$), the optimal cell association for the uplink is shown.}
  \label{fig:Nctwo}
\end{figure}

In this section, we explain the answer to the considered problem through the simple scenario when each mobile terminal can be associated with two base stations. The optimal downlink cell association for this case is illustrated in Figure~\ref{fig:nctwodownlink}. For the uplink, the optimal cell association in this case is simply associating each mobile terminal with the two base stations connected to it, and this guarantees achieving one degree of freedom per user as follows. The last base station decodes the last message and then passes it to base station $K-1$. Starting with base station $K-1$ and moving in the direction of decrementing the base station index, each base station will decode the message with the same index and then pass it to the base station with a previous index. The optimal cell association for the uplink is illustrated in Figure~\ref{fig:nctwouplink}. 

Now consider the average per user DoF achieved over both the uplink and downlink sessions. The cell association of Figure~\ref{fig:nctwodownlink} enables achieving a downlink per user DoF of $\frac{4}{5}$, but the maximum per user DoF that can be achieved using this same cell association is also $\frac{4}{5}$ (proof will follow in Section~\ref{sec:converse}), yielding an average per user DoF of $\frac{4}{5}$. The cell association depicted in Figure~\ref{fig:nctwouplink} enables achieving an uplink per user DoF of unity, while the maximum achievable downlink per user DoF is $\frac{2}{3}$ (see~\cite{Ntranos-uplink} for the proof), yielding a higher average of $\frac{5}{6}$ per user DoF. 

We can observe through the example of Figure~\ref{fig:nctwouplink}, that using the cell association that is optimal for the uplink, an average per user DoF of $\frac{5}{6}$ can be achieved for the case when $N_c=2$. In general, this value can be generalized to $\frac{4 N_c -3}{4 N_c - 2}$, and we can prove that this is the best achievable average per user DoF for any choice of the cell association that satisfies the backhaul constraint of~\eqref{eq:backhaul_constraint}. 

\section{Main Result}\label{sec:main}
We characterize the average per user DoF over both uplink and downlink sessions in Theorem~\ref{thm:main} below.
\begin{thm}\label{thm:main}
For the linear interference network described in Section~\ref{sec:channel} and the backhaul constraint of~\eqref{eq:backhaul_constraint}, the average per user DoF is given by,
\begin{equation}\label{eq:main}
\tau(N_c)=\begin{cases}
\frac{2}{3}, & N_c=1,\\
\frac{4N_c -3}{4N_c - 2}, & N_c \geq 2.
\end{cases}
\end{equation}
\end{thm}

We observe from~\eqref{eq:main} that for every value of $N_c \geq 2$, the average per user DoF is higher than the per user DoF for the downlink only, since in~\cite{ElGamal-Annapureddy-Veeravalli-IT14}, it is shown that $\tau_D(N_c)=\frac{2N_c}{2N_c+1}$. The reason is that through message passing between base station receivers in the uplink, interference can be eliminated if all the base stations connected to a mobile terminal are associated with it, and this is possible when $N_c \geq 2$, and hence, one degree of freedom per user can be achieved. Further, as we will explain in detail in Section~\ref{sec:achievability}, the average per user DoF when $N_c \geq 2$, is an average of a unity uplink per user DoF  and the downlink per user DoF with reduced backhaul capacity. More precisely,

\begin{equation}
\tau(N_c) = \frac{1+\tau_D(N_c-1)}{2}, \forall N_c \geq 2.
\end{equation}


We provide the proof of the lower and upper bounds of Theorem~\ref{thm:main} in Sections~\ref{sec:achievability} and~\ref{sec:converse}. We find the converse proof of this theorem interesting and hope that deeper understanding of the argument can help us understand in general for which network topologies we would get the same answer. That is, the optimal cell association would guarantee achievability of the best uplink rate if the backhaul constraint is greater than or equal to the number of base stations connected to a mobile terminal.

\section{Proof of Achievability}\label{sec:achievability}
For the case where $N_c=1$, we divide the network into subnetworks; each consisting of successive three indices. In each subnetwork, the mobile terminal with the middle index is deactivated and each other mobile terminal enjoys interference-free communication in both the uplink and downlink sessions. The cell association enabling interference-free communication is as follows.

\begin{equation}\label{eq:ncone}
{\cal C}_i = \begin{cases}
\{i\}, & i \text{ mod } 3 = 1, \\
\{i-1\}, & i \text{ mod } 3 = 0.
\end{cases}
\end{equation}

It follows that $\frac{2}{3}$ per user DoF is achieved in each of the downlink and uplink sessions. For the case where $N_c = 2$, we associate each mobile terminal with the two base stations connected to it. More precisely,
\begin{equation}\label{eq:ncgeqtwo}
{\cal C}_i= \begin{cases} 
\{1\}, & i=1, \\
\{i,i-1\}, & i \geq 2.
\end{cases}
\end{equation}

For the uplink, one degree of freedom per user can be achieved since for $1 \leq i \leq K-1$, base station $i$ receives the decoded message $M_{i+1}$ from base station $i+1$, and hence, with the availability of the channel state information at the base stations, it can cancel the interference due to message $M_{i+1}$ and decode message $M_i$ without interference. 

For the downlink, $\frac{2}{3}$ per user DoF is achieved by dividing the network into subnetworks; each consists of three indices, and the last base station transmitter and mobile terminal receiver in each subnetwork are both deactivated. The first mobile terminal in each subnetwork gets interference-free communication because the last base station transmitter in the preceding subnetwork is inactive. The second base station transmitter in each subnetwork uses its knowledge of the first message to cancel its interference at the second mobile terminal receiver. Hence, two mobile terminal receivers in each subnetwork have interference-free communication. The average achieved per user DoF for $N_c=2$ is hence $\frac{1+\frac{2}{3}}{2}=\frac{5}{6}$, which is the value stated in~\eqref{eq:main}. 

For $N_c > 2$, we first associate each mobile terminal with the two base stations connected to it, and hence, one per user degree of freedom is achieved in the uplink as in the case of $N_c=2$. For the downlink, we now explain how to achieve $\frac{2N_c-2}{2N_c-1}$ per user DoF, yielding an average per user DoF of $\frac{4N_c-3}{4N_c-2}$. Recall that each mobile terminal is already associated with the two base stations connected to it, we assign the remaining $N_c-2$ associations by using the downlink scheme described in~\cite{ElGamal-Annapureddy-Veeravalli-IT14} for the case when each message can be available at $N_c-1$ base station transmitters, and note that this is possible since in the optimal scheme of~\cite{ElGamal-Annapureddy-Veeravalli-IT14}, each active mobile terminal is associated with at least one of the base stations connected to it. More specifically, the network is split into subnetworks; each consists of successive $2N_c-1$ indices. The middle mobile terminal receiver and the last base station transmitter in each subnetwork are inactive. We explain the cell associations for the first subnetwork and the rest will follow in a similar fashion. Define the following two subsets of indices in the first subnetwork: ${\cal S}_1=\{1,2,\cdots,N_c-1\}, {\cal S}_2=\{N_c+1,N_c+2,\cdots,2N_c-1\}$, then the cell associations are determined as follows.

\begin{equation}\label{eq:ncgtwo}
{\cal C}_i = \{i-1,i\} \cup \begin{cases}
\{i,i+1,\cdots,N_c-1\}, & i\in {\cal S}_1, \\
\{i-1,i-2,\cdots,N_c\}, & i\in {\cal S}_2.
\end{cases}
\end{equation} 

The proof that the cell associations described in the cases listed in~\eqref{eq:ncgtwo} enable the achievability of $2N_c-2$ degrees of freedom in each subnetwork follows from~\cite{ElGamal-Annapureddy-Veeravalli-IT14}.
\section{Converse Proof}\label{sec:converse}
For the case where $N_c=1$, the upper bound follows from the fact that the maximum per user DoF for each of the downlink and uplink sessions is $\frac{2}{3}$, even if we are allowed to change the cell association between the uplink and downlink. The proof of the downlink case is provided in~\cite{ElGamal-Annapureddy-Veeravalli-IT14}. The proof of the uplink case is similar to the downlink case, so we omit it here for brevity and instead focus in the rest of the section on the remaining and more difficult case of $N_c \geq 2$. 

Before making the main argument, we first need the following auxiliary lemmas for finding a converse for the uplink scenario.

\begin{lem}\label{lem:uplinklemone}
Given any cell association and any coding scheme for the uplink, the per user DoF cannot be increased by adding an extra association of mobile terminal $i$ to base station $j$, where $j \notin \{i,i-1\}$.
\end{lem}
\begin{IEEEproof}
The lemma states that associating any mobile terminal to a base station that is not connected to it cannot be \emph{useful} for the uplink case. The key fact validating this lemma is that unlike the downlink case, the knowledge of a message at a base station cannot allow for the possibility of propagating the interference caused by this message beyond the two original receivers that are connected to the transmitter responsible for delivering the message. In other words, no matter what cell association we use for mobile terminal $i$, the message $M_i$ will not cause interference at any base station except base station $i$ and $i-1$, and hence, having this message at any other base station cannot help neither in decoding the message nor in canceling interference. We omit the formal information-theoretic proof here for brevity.
\end{IEEEproof}

Lemma~\ref{lem:uplinklemone} gives us two possibilities for choosing the cell association of mobile terminal $i$; either we associate it with both base stations $i$ and $i-1$ or only one of these base stations. We use the following lemma to upper bound the degrees of freedom for the latter case.

\begin{lem}\label{lem:uplinklemtwo}
If either mobile terminal $i$ or mobile terminal $i+1$ is not associated with base station $i$, then it is either the case that the received signal $Y_i$ can be ignored in the uplink without affecting the sum rate, or it is the case that the uplink sum DoF for messages $M_i$ and $M_{i+1}$ is at most one.
\end{lem}
\begin{IEEEproof}
If neither $M_i$ nor $M_{i+1}$ is associated with base station $i$, then it is clear that $Y_i$ can be ignored in the uplink. Further if only one of the two message is associated with base station $i$ but is not decodable from $Y_i$ in the uplink, then we also can ignore this received signal. We now focus on the reamaining case when exactly one of $M_i$ and $M_{i+1}$ is associated with base station $i$ and can be successfully decoded from $Y_i$ in the uplink. The proof is similar to the proof of~\cite[Lemma $5$]{ElGamal-Annapureddy-Veeravalli-IT14}; we only provide a sketch here. First assume without loss of generality that message $M_i$ is the message associated with base station $i$. Assuming a reliable communication scheme, if we are given the received signal $Y_i$, message $M_i$ can be decoded reliably, and hence, the transmit signal $X_i$ can be reconstructed. Since the channel state information is available at base station $i$, and $Y_i$ only depends on $X_i$ and $X_{i+1}$ and $Z_i$, then the remaining uncertainty in reconstructing the signal $X_{i+1}$ is only due to the Gaussian noise. Since the uncertainty in the Gaussian noise does not reduce the degrees of freedom, we ignore it in our argument. Now, $M_{i+1}$ can be recovered since we know $X_{i+1}$. Since we could recover both $M_i$ and $M_{i+1}$ using only one received signal, it follows that $d_i + d_{i+1} \leq 1$. 
\end{IEEEproof}

We first consider the case where each mobile terminal can be associated with two base stations, i.e., $N_c=2$, and only provide a sketch of the argument here because of the page limit. Fix a cell association and divide the indices of the network into sets; each consists of consecutive three indices. Note that we only care about the ratio of the degrees of freedom to the number of users in large networks, so there is no loss in generality in assuming that $K$ is a multiple of $3$. For each subnetwork, if the middle base station is only associated with at most one of the mobile terminals that are connected to it, then it follows from Lemma~\ref{lem:uplinklemtwo} that the uplink per user DoF for users in the subnetwork is at most $\frac{2}{3}$. If the middle base station is associated with more than two mobile terminals connected to it, then we can show for the downlink, that given only the second and third received signal in this subnetwork, one can reconstruct all three transmit signals; this can be proven in this case by using an argument similar to the proof of~\cite[Lemma $4$]{ElGamal-Annapureddy-Veeravalli-IT14}, since given the second and third received signals, we can reconstruct the middle transmit signal, and then from the connectivity of the linear interference network, we can reconstruct the third transmit signal from the third received signal, and the first transmit signal from the second received signal. 

We now make the main argument to prove the converse for $N_c=2$. In order to achieve an average per user DoF that is greater than $\frac{5}{6}$, the uplink per user DoF has to exceed $\frac{2}{3}$. Assume that the uplink per user DoF for a given cell association is $x$, where $x \geq \frac{2}{3}$, and divide the network into subnetworks, each consisting of three consecutive indices. It follows from the argument in the previous paragraph that for at least a fraction of $(3x-2)$ of the subnetworks, the middle base station is associated with both the second and third mobile terminals. It hence follows that for a fraction of $(3x-2)$ subnetworks in the downlink, two mobile terminal received signals suffice to reconstruct all three base station transmit signals. We can then reconstruct all transmit signals in the downlink from a number of transmit and received signals that equals a fraction of the number of users that equals $\frac{2}{3}(3x-2)+1-(3x-2)=\frac{5}{3}-x$ for large networks. The per user DoF in the downlink is then at most $\frac{5}{3}-x$, and hence, the average per user DoF is at most $\frac{5}{6}$. The formal information-theoretic argument used to prove the downlink upper bound will be provided in the journal version of this work and has similarity to the proof of~\cite[Lemma $4$]{ElGamal-Annapureddy-Veeravalli-IT14}.

We now generalize the above argument to prove the converse for the case where $N_c \geq 2$. For the average per user DoF to exceed $\frac{4N_c-3}{4N_c-2}$, the uplink per user DoF has to exceed $\frac{2N_c-2}{2N_c-1}$. Fix a cell association for which the uplink per user DoF equals $x$, where $x \geq \frac{2N_c-2}{2N_c-1}$, and divide the network into subnetworks, each consisting of $2N_c-1$ successive indices. If the middle base station in each subnetwork is associated with at most one of the two mobile terminals connected to it, then Lemma~\ref{lem:uplinklemtwo} implies that the uplink per user DoF in this subnetwork is at most $\frac{2N_c-2}{2N_c-1}$. It follows that for at least a fraction of $(2N_c-1)x-(2N_c-2)$ of the subnetworks, the middle base station is associated with the two mobile terminals connected to it. For these subnetworks, we can show that the knowledge of $2N_c-2$ received signals suffices to reconstruct all $2N_c-1$ transmit signals, and hence, the per user DoF in the downlink is at most $\frac{4N_c-3}{2N_c-1}-x$. It follows that the average per user DoF is at most $\frac{4N_c-3}{4N_c-2}$.

\section{Concluding Remarks and Future Directions}\label{sec:conclusion}
The obtained result in this work indicates that the optimal cell association depends on the number of base stations that each mobile terminal can be associated with ($N_c$). If $N_c$ is less than the number of base stations connected to each mobile terminal then the downlink-optimal association is average-optimal; otherwise, the optimal cell association first assigns each mobile terminal to the two base stations connected to it to guarantee achieving the optimal uplink rate, and then assigns the remaining $N_c-1$ base stations according to a downlink-optimal strategy with a reduced backhaul constraint. The main question we have for future work is whether this insight holds for more general and practically relevant cellular network models than the considered linear interference model, as for example the cellular models considered in~\cite{cellular}.

The other comment we have is for the dynamic interference network model introduced in~\cite{dynamic}, where each link between a mobile terminal and a base station can be erased with a fixed probability in any given block of communication sessions. It was shown in~\cite{dynamic} that for the downlink, there is no \emph{universally optimal} cell association over all values of the erasure probability. However, the considered setting might change this conclusion for dynamic networks. For example, if we consider the uplink only, assigning each mobile terminal to the two base stations connected to it is universally optimal for any value of $N_c \geq 2$. The question we pose is when optimizing for the average over both uplink and downlink, what result would we get for dynamic interference networks?

\bibliographystyle{IEEEtran}

\end{document}